\documentclass[journal]{IEEEtran}
\usepackage{cite}

\usepackage{graphicx}
\ifCLASSINFOpdf
\else
\fi
\def\BibTeX{{\rm B\kern-.05em{\sc i\kern-.025em b}\kern-.08em
    T\kern-.1667em\lower.7ex\hbox{E}\kern-.125emX}}
\usepackage{amsmath,amssymb,amsfonts}
\usepackage{mathrsfs}
\newtheorem{theorem}{Theorem}
\newtheorem{lemma}{Lemma}

\newtheorem{problem}{Problem}
\newtheorem{assume}{Assumption}
\newtheorem{remark}{Remark}

\newcommand{\Rn}{\mathbb{R}^n}

\usepackage{algorithm,algorithmic}
\usepackage{mathtools}
\usepackage{comment}

\usepackage{array}
\usepackage{xcolor}

\begin{document}
%
\title{ADMM based Distributed State Observer Design under Sparse Sensor Attacks}
%
%
%
\author{Vinaya~Mary~Prinse
        and~Rachel~Kalpana~Kalaimani
\thanks{The authors are with the Department of Electrical Engineering, Indian Institute of Technology Madras, Chennai 600036 India (email: ee20s021@smail.iitm.ac.in, rachel@ee.iitm.ac.in).}}
\maketitle

\begin{abstract}
This paper considers the design of a distributed state-observer for discrete-time Linear Time Invariant (LTI) systems in the presence of sensor attacks. We assume there is a network of observer nodes, communicating with each other over an undirected graph, each with partial measurements of the  output corrupted by some adversarial attack. 
We address the case of sparse attacks where the attacker targets a small subset of sensors.  
An algorithm based on Alternating Direction Method of Multipliers (ADMM) is developed which provides an update law for each observer which ensures convergence of each observer node to the actual state asymptotically.
\end{abstract}


%
\IEEEpeerreviewmaketitle

\section{Introduction}
%
%
%
%
\IEEEPARstart{O}{}ver the past few decades, a distributed approach is being adopted for large scale and complex Cyber-Physical Systems (CPSs) such as smart grids, industrial control systems, robotic systems etc. to enhance flexibility, robustness and computational performance. However, the susceptibility of these systems to attacks is a reality as highlighted in \cite{10.5555/1496671.1496677} and \cite{article} and a few examples include the Ukrainian power grid hack\cite{7592621}, the worldwide Wannacry ransomware attack and the Stuxnet attack \cite{9763485}. Hence, the security of CPSs is of primary concern. 

We consider the problem of estimating  the state of linear dynamical systems when few sensors are corrupted, called sparse sensor attacks. We adopt a distributed approach for an observer design and assume a network of observers where each observer has access to partial measurements of the output. Since each observer need not be observable, state estimation by the observers require communication with each other. 

A {\em centralized} state observer for Linear Time Invariant (LTI) systems under sparse sensor attacks is proposed in \cite{7308014} and \cite{LU2018124} considers sparse actuator attacks also. Design of a \emph{distributed} observer for state estimation \emph{without sensor attacks} is discussed in \cite{8270595} for discrete time LTI systems and in \cite{8340829} for continuous time LTI systems. 

While an observer is a dynamic process, which uses new measurements to update the current state value, a static approach is to just estimate the initial condition using a batch of measurements and then use the system dynamics to construct the current state. Now, additionally when there are sensor attacks, estimating the state using the above static approach is known as Secure State Estimation (SSE) problem in literature.  When a distributed approach is used, the above problem is referred to as Distributed SSE (DSSE) where  each agent estimates the \emph{initial state} based on its own (limited) state measurements and the information from neighbours despite sparse sensor attacks. DSSE has been addressed in \cite{AN2019526} and \cite{07654887318044adab44fdddb1abbca3}. The latter has an additional sparsity assumption on the initial state. 
DSSE with secure preselectors is discussed in \cite{8580572} but has a complicated parameter design as mentioned in \cite{9763485}. 

 A distributed observer for LTI systems with Byzantine attacks is studied in \cite{Mitra2019ByzantineresilientDO}. In Byzantine attacks, adversaries are allowed to send differing state estimates to different neighbors at the same instant of time,  State estimation for this kind of adversaries is addressed for
 stronger assumptions on the network graph. 
We consider a simpler case, where the attacker tampers only a few measurements of the system and therefore require a comparatively  weaker assumption of just the graph being connected. 
 
A distributed observer design in case of sparse attacks is proposed in \cite{8619126} for linear continuous time systems. This  involves an attack indicator signal with exponential computational complexity. We propose a distributed observer design, for discrete time LTI systems, under sparse sensor attacks.  The main contributions of our work are as follows:
\begin{enumerate}
    \item Design a  distributed observer when there are sparse sensor attacks. An algorithm based on 
     Alternating Direction Method of Multipliers (ADMM) is used to update  each observer. 
    \item The DSSE can be recovered as a special case of the above algorithm. We compare this with the DSSE algorithm in \cite{07654887318044adab44fdddb1abbca3}.
    \item We also compare the centralized implementation of our algorithm 
   with \cite{7308014} which discusses a centralized state observer under sparse sensor attacks.
\end{enumerate}
\subsection{Notation} 
$\mathbb{N,R}$ denote the set of natural and real numbers respectively. $\bar{\Gamma}$ denotes complement of a set $\Gamma$. $b^T$ represents transpose of vector $b$. $I_{n}$ is the identity matrix of dimension $n$. For a vector $y\in\mathbb{R}^n$, $l_0$ norm of $y$ i.e. $\|y\|_0$ refers to the number of nonzero components in $y$ and $\|y\|_r$ represents r norm of $y$. $(x,y)$ denotes the vector $[x^T\hspace{1.2ex}y^T]^T\subset\mathbb{R}^{n_1}\times\mathbb{R}^{n_2}$ where $x\in\mathbb{R}^{n_1}$ and $y\in\mathbb{R}^{n_2}$. 

$|K|$ represents cardinality of a set $K$. 
A vector $x\in\Rn$ is said to be $s$ sparse when $|supp(x)|\leq s$. A block vector given by $y=(y_1,...,y_p)\in\mathbb{R}^{p\tau}$, where $y_1,...,y_p\in\mathbb{R}^\tau$ are blocks, is block $s$-sparse if at most $s$ blocks are nonzero. We use $s$-sparse instead of the term block $s$-sparse and $\mathbb{S}_s^{p\tau}\subset\mathbb{R}^{p\tau}$ denotes the set of these vectors.

An undirected graph is represented by $\mathcal{G}=(\mathcal{V,E,A})$ where $\mathcal{V}=\{1,2,..,N\}$ is the vertex set, $\mathcal{E}\subset\mathcal{V}\times\mathcal{V}$ is the undirected edge set and $\mathcal{A}=\{a_{ij}\}_{N\times N}$ is the adjacency matrix where $a_{ij}=1$ if $(i,j)\in \mathcal{E}$ i.e. it indicates that nodes $i$ and $j$ are connected, else $a_{ij}=0$. We will assume a graph without loops or multiple edges i.e.  $a_{ii}=0$. The neighbourhood set of the $i$th node is defined as $\mathbb{N}_i=\{j:(i,j)\in \mathcal{E}\}$. A graph is said to be connected if there exists a path to traverse between every pair of distinct nodes $i$ and $j$ in $\mathcal{G}$. 

Hereafter the paper organization is as follows. 
Section \ref{prob} formally states the problem objective. The main results of this paper are given in Section \ref{opt} which includes optimization problem reformulation followed by the distributed observer design and simulation results for DSSE and centralized observer implementations in addition. The paper is concluded in section \ref{conc}.
\section{Problem Formulation} \label{prob}
Consider the following discrete-time LTI system, whose measurements are corrupted.
\begin{eqnarray}\label{eqn:sys_dynamics}
x[t+1]&=&Ax[t]+Bu[t],\nonumber \\
y[t]&=&Cx[t]+a[t],
\end{eqnarray}
where $x[t]\in\mathbb{R}^{n}$ is the state vector, $y[t]\in\mathbb{R}^{p}$ is the measurement vector and $a[t]\in\mathbb{R}^{p}$ is the $s$ sparse attack vector assuming that at most $s$ sensors are attacked.
The objective is to construct a distributed observer for the above system which estimates the system state and attack vector (thereby the set of corrupted sensors) at time $t$ by communication within the network.

It is well known that a sufficient condition which guarantees exact state estimation despite an $s$ sparse sensor attack is the $2s$-sparse observability condition \cite{7308014}. A system is said to be $s$-sparse observable if the system remains observable even after removing any $s$ sensors i.e. for every set $\Gamma\subset\{1,..,p\}$ with $|\Gamma|=s$, the observability matrix of $(A,C_{\bar{\Gamma}})$ has full rank ($C_{\bar{\Gamma}}$ indicates the matrix resulting on elimination of rows indexed by $\Gamma$ from $C$). 

For the distributed framework, we assume that there are $N$ local observers (or agents) each with a set of sensors which measures a  part of the output, $y_i[t]\in\mathbb{R}^{p_i}$, $\sum_{i=1}^N p_i=p$. This measurement is corrupted by $a_i[t]\in\mathbb{R}^{p_i}$. The communication network is depicted by a graph, whose vertices represent the observers and edges represent the existence of a communication channel between two observers. Using the partial, corrupted measurement and additionally communicating with its neighbours, each observer should be able to estimate the state vector of the system in \eqref{eqn:sys_dynamics}. 
 
Next we list the assumptions on our system model:
\begin{assume}
\begin{enumerate}
    \item The graph representing the communication network of the distributed observers is connected.
    \item The system $(A,C)$ is $2s$-sparse observable
    \item The attack vector $a[t]$ is $s$ sparse (provided $s<\frac{p}{2}$ \cite{6120187})
\end{enumerate}
\end{assume}
The set of corrupted sensors is assumed to be fixed over time. However, it is important to note that the effect of the attack is propagated to the neighboring observers of a local observer receiving corrupted sensor measurements, through the communication network.
Next we formally state our problem of designing a distributed observer under sensor attacks.
\begin{problem}Consider a set of $N$ agents/observers, each with partial and corrupted measurements of the output, interacting with each other to estimate the state of the system in \eqref{eqn:sys_dynamics}, satisfying conditions in Assumption 1.
Let $\tilde{x_i}[t]\in\mathbb{R}^n$ denote an estimate of the state
that an observer $i$ has at time $t$. Design an update law for each agent, that can be implemented in a distributed manner, such that the following holds:
\[\underset{t\rightarrow \infty}{\mbox{lim}}\|\tilde{x_i}[t]-x[t]\|\rightarrow 0, ~\forall~ i\]
\end{problem}
\section{Results} \label{opt}
\subsection{Optimization Formulation}
In \cite{7308014}, in order to obtain an observer, 
an optimization problem is formulated first and then recursively solved. We  follow a similar approach to get a {\em distributed} observer. By collecting the sequence of last $\tau$ observed outputs ($\tau\in\mathbb{N},\tau\leq n$) at time $t\geq \tau $, a delayed version of the system state i.e. $x[t-\tau+1]$ can be reconstructed.

Let $E_j[t]\in\mathbb{R}^\tau$ denote the vector of attack signals on the $j$th sensor from time $t-\tau+1$ to $t$, i.e. $E_j[t]=(a_j[t-\tau+1],...,a_j[t])$.  This vector is stacked for all sensors to obtain  $E[t]$ i.e. $E[t]=(E_1[t],...,E_p[t])$.
The attack signal on the $j$th sensor, $a_j[t]$, can be expressed as 
 \begin{equation}\label{eqn:attack_vec}
  a_j[t]=y_j[t]-C_jx[t]  
 \end{equation}
 Hence, if we have an estimate of $x[t-\tau+1]$, then using the
 above equation \eqref{eqn:attack_vec} and the system dynamics in \eqref{eqn:sys_dynamics}, we can obtain an estimate of the state and attack vectors from time $t-\tau+1$ to $t$. 
 
Consider the vector $z(t)$ defined as
 $z[t]=(x[t],E[t])$. Let $\bar{Y}(t)\in\mathbb{R}^{p\tau}$
denote the set of past $\tau$ outputs collected for all the sensors i.e. $\bar{Y}[t]=(\bar{Y}_1[t],..., \bar{Y}_p[t])$ where $\bar{Y}_j[t]=( y_j[t-\tau+1],...,y_j[t])$. Then from \cite{7308014}, we obtain the dynamics of $z(t)$ and a relation between $\bar{Y}(t)$ and $z(t)$ as follows: 
\begin{eqnarray}
z[t]&=&\bar{A}z[t-1]+Ny[t] \label{eqn:dyn_z}\\
\bar{Y}[t]&=&Qz[t]\label{eqn:Yandz}
\end{eqnarray}
where
$\bar{A}=\left[\begin{smallmatrix}
     A & 0 & \cdots{} & 0  \\
     G_1 & S & \cdots{} & 0 \\
     \vdots{}& &\ddots{} & \\
     G_p & 0 & \cdots{} & S
\end{smallmatrix}\right]$, $~Q=\left[\begin{smallmatrix}
O_1  & \\
\vdots{}  & I_{p\tau} \\
O_p & 
\end{smallmatrix}\right]~$ and $~N=\left[\begin{smallmatrix}
0 \\ N_1 \\ \vdots{} \\ N_{p}
\end{smallmatrix}\right]$. 
Here,
$G_j=\left[\begin{smallmatrix}
0\\ \vdots{} \\ 0 \\ -C_jA^\tau
\end{smallmatrix}\right]$, $S=\left[\begin{smallmatrix}
0&1&\cdots{}&\cdots{}&0\\
0&0&1&\cdots{}&0\\
\vdots{}&\vdots{}& &\ddots{}& \\
0&0&\cdots{}&\cdots{}&1\\
0&0&\cdots{}&\cdots{}&0
\end{smallmatrix}\right]$ and $N_j=\left[\begin{smallmatrix}
0\\ \vdots{} \\ 0\\b_j
\end{smallmatrix}\right]$ where $b_j$ represents the $j$th standard basis vector and $O_j=\left[\begin{array}{cccc}
C_j^T & (C_jA)^T & \hdots{} & (C_jA^{\tau-1})^T
\end{array}\right]^T$.

Since we are interested in an observer, for simplifying the analysis, we ignore the external input in the system dynamics. Let $\tilde{x}[t]$ denote the estimate of $x[t-\tau+1]$ and  $\tilde{E}[t]$
denote the estimate of $E[t]$. Then estimate of $z[t]$ denoted as $\tilde{z}[t]$ is the vector $(
\tilde{x}[t], \tilde{E}[t])$.

In order to obtain the estimate $\tilde{z}(t)$ at a given time $t$, the following optimization problem is solved:
\begin{equation}\label{cent_optprob}
   \underset{\tilde{z}\in \mathbb{R}^{n}\times\mathbb{S}_s^{p\tau}}{\min}\frac{1}{2}\|\bar{Y}[t]-Q\tilde{z}[t]\|_2^2
\end{equation}
Note that $\tilde{E}$ part of the variable $\tilde{z}$ in the above optimization problem is constrained to be in $\mathbb{S}_s^{p\tau}$ due to the sparse structure of the attack vector. According to Theorem 3.2 in  \cite{7308014}, for a $2s$-sparse observable system, this problem has a unique minimum, $(x^*,E^*)$ in the set $\mathbb{R}^{n}\times\mathbb{S}_s^{p\tau}$.
Hence, this problem is reformulated as an $l_0$ norm minimization problem for $\tilde{E}$. The optimal solution, which is the \textit{most sparse} $\tilde{E}$ will be the unique $s$-sparse vector $E^*$ in this case.
Thus, the sparse nature of the attack vector is utilised to reformulate the above problem into the following $l_0$ norm optimization problem $A$:
\begin{equation} \label{zero norm prob}
     A: \underset{\tilde{z}=(\tilde{x},\tilde{E})}{\min}\|\tilde{E}\|_0 \hspace*{1ex} s.t. \hspace*{1ex} Q\tilde{z}=\bar{Y}   
\end{equation}

Note that problem $A$ is not convex. Hence we consider the following convex relaxation of problem $A$ known as the basis pursuit problem:
\begin{equation} \label{E 1 norm prob}
     B: \underset{\tilde{z}=(\tilde{x},\tilde{E})}{\min}\|\tilde{E}\|_1 \hspace*{1ex} s.t. \hspace*{1ex} Q\tilde{z}=\bar{Y}   
\end{equation}
The following lemma provides a condition when the two problems are equivalent. \cite[Proposition~6]{6120187} and \cite[Proposition~3]{8664617} discuss similar conditions and hence we skip the proof.
\begin{lemma}\label{lem:convex_relaxn} 
Let $(\tilde{x}_0,\tilde{E}_0)$ and $(\tilde{x}_1,\tilde{E}_1)$ be the solutions to the optimization problems $A$ and $B$ respectively. Then for a $2s$-sparse observable system, the following are equivalent:
\begin{enumerate}
    \item $(\tilde{x}_1,\tilde{E}_1)=(\tilde{x}_0,\tilde{E}_0)$
    \item For all $\Gamma \subset \{1,...,p\}$ with $|\Gamma|=s$, the following holds:
\begin{equation} \label{convex_relaxn_condn}
   \sum_{i\in\Gamma} |(O\tilde{x})_i| < \sum_{i\in\bar{\Gamma}}|(O\tilde{x})_i|, ~\forall~ \tilde{x}\in\mathbb{R}^n\symbol{92}\{0\}  
\end{equation}
where 
$O=[O_1^T  ~~\cdots{}~~ O_p^T]^T$ is part of the $Q$ in \eqref{eqn:Yandz}.
\end{enumerate}
\end{lemma}
These results are utilised to reformulate the optimization problem in \eqref{cent_optprob} into a basis pursuit problem in the following theorem.

\begin{theorem}
Let the discrete time LTI system defined in \eqref{eqn:sys_dynamics} be $2s$-sparse observable.  Then the optimization problems in \eqref{cent_optprob} and \eqref{E 1 norm prob}
are equivalent provided condition \eqref{convex_relaxn_condn} holds $\forall~\Gamma\subset\{1,...,p\}$ with $|\Gamma|=s$.
\end{theorem}
The proof is straightforward from the re-formulated optimization problem A in \eqref{zero norm prob} and Lemma \ref{lem:convex_relaxn}.
$\hfill \blacksquare$

Problem $B$ needs to be formulated in a distributed set up in order to address our main objective of proposing a distributed observer. Each local observer in the network will have access to a part of the measurements of the system and collection of these measurements from time $t-\tau+1$ to $t$ gives $Y_i[t]$ for the $i$th observer. Now an observer/agent $i$ can have an estimate of the state vector $x[t-\tau+1]$ and of the attack vector corresponding to the measurements received by them denoted by $\tilde{x_i}[t]$ and $\tilde{\mathbb{E}_i}[t]$ respectively.
 Similar to the output equation given in \eqref{eqn:Yandz}, the output equation for observer $i$ can be written as follows:
    \begin{equation} \label{agent output equation}
    \begin{split}
    Y_i[t]&=
    \left[\begin{array}{cc}
O_{1+\sum_{j=1}^{i-1}p_j} & \\
\vdots{} & I_{p_i\tau} \\
O_{p_i+\sum_{j=1}^{i-1}p_j} & 
\end{array}\right]\left[\begin{array}{c}
\tilde{x_i}[t]\\
\tilde{\mathbb{E}}_i[t]
\end{array}\right]=Q_i\tilde{z_i}[t]
    \end{split}
\end{equation}
where $\tilde{\mathbb{E}}_i[t]$ is a vector extracted from $\tilde{E}[t]$ corresponding to the attack on observer i and $Q_i$ refers to the matrix in the above equation.
Each agent gets an  estimate of the initial condition and attack vectors, $\tilde{z_i}[t]=(\tilde{x_i}[t],\tilde{\mathbb{E}_i}[t])$, by a distributed approach. In this regard, an optimization problem that is equivalent to $B$ is formulated in  the following Lemma.
\begin{lemma} \label{lem:dist.prob_reform}
The optimization problem $B$ in \eqref{E 1 norm prob} is equivalent to the following optimization problem: 
\begin{align}\label{dist_SBO}
 \underset{\tilde{z_i}=(\tilde{x_i},\tilde{\mathbb{E}}_i),b_i}{\min}\|\tilde{\mathbb{E}}_1\|_1 + \|\tilde{\mathbb{E}}_2\|_1 + ... + \|\tilde{\mathbb{E}}_N\|_1 \\ s.t. \hspace*{1ex} Q_{i}\tilde{z_i}=Y_i \nonumber \hspace*{1ex} \forall i=1,...,N\\ \tilde{x_i}=b_j, \nonumber \hspace*{1ex} \forall (i,j) \in \mathcal{E}\\ \tilde{x_i}=b_i, \hspace*{1ex} \forall i = 1,...,N \nonumber
 \end{align}
where $Q_i$ is defined in \eqref{agent output equation} and $b_i$ is an auxiliary variable. 
\end{lemma}

\textit{Proof:} In problem $B$, the objective function $\|\tilde{E}\|_1$ can be written as the sum of $\|\tilde{\mathbb{E}}_i\|_1$s by definition of one norm.
The constraint $Q\tilde{z}=\bar{Y}$, is decomposed for each observer $i$ in \eqref{agent output equation}. 
Consequently, each observer maintains its own estimate of the initial condition, $\tilde{x_i}$. 
Next, to enforce consensus on the state estimates i.e. to achieve $\tilde{x_1}=...=\tilde{x_N}$ in a distributed/parallel manner, we need to introduce auxiliary variables. We adopt the approach proposed in \cite{6365874} which uses one auxiliary variable per node, say $b_i$. Then the consensus constraint equations can be written as $\tilde{x_i}=b_j \hspace*{1ex} \forall (i,j) \in \mathcal{E}$ and $\tilde{x_i}=b_i \hspace*{1ex} \forall i = 1,...,N$. Since the objective function and constraints are equivalent, the optimization problems given in \eqref{E 1 norm prob} and \eqref{dist_SBO} are equivalent.
$\hfill \blacksquare$

\subsection{Distributed Observer}\label{subsec:distr_observer}
In this section, we first propose an ADMM based algorithm, that can be implemented in a distributed manner, to solve the optimization problem formulated in Lemma \ref{lem:dist.prob_reform}. Then for the observer, a suitable recursive implementation is proposed incorporating new measurements at each time.

Consider the augmented Lagrangian for the optimization problem given in \eqref{dist_SBO} (by dualizing only the consensus constraints) :
\begin{multline*}
     \mathcal{L}_\rho=\sum_{i=1}^{N}(\|\tilde{E_i}\|_1+\sum_{j\in\mathcal{N}_i}\lambda_{ij}^T(\tilde{x_i}-b_j)+\frac{\rho}{2}\sum_{j\in\mathcal{N}_i}\|\tilde{x_i}-b_j\|^2 \\+\lambda_{ii}^T(\tilde{x_i}-b_i)+\frac{\rho}{2}\|\tilde{x_i}-b_i\|^2)
 \end{multline*}
where $\lambda_{ij},\lambda_{ii}$ $\in \mathbb{R}^n$ are the Lagrange multipliers associated with the constraints $\tilde{x_{i}}=b_j$ and $\tilde{x_{i}}=b_i$ respectively. The primal and dual update steps for each agent are as follows:
\begin{multline*}
\tilde{z_{i}}^{k}= \underset{\tilde{z_{i}}=(\tilde{x_i},\tilde{E_i})}{\arg\min}\|\tilde{E_i}\|_1 + \sum_{j\in \mathcal{N}_i}(\lambda_{ij}^{k-1}+\lambda_{ii}^{k-1})^T \tilde{x_{i}} + \\ \frac{\rho_i^{k}}{2}\sum_{j\in \mathcal{N}_i \cup {i}}\|\tilde{x_{i}}-b_{j}^{k-1}\|^2 \hspace*{1ex}
s.t. \hspace*{1ex} Q_{i}\tilde{z_i}=Y_i    
\end{multline*}
\begin{equation*}
b_{i}^{k}= \underset{b_{i}}{\arg\min}\hspace*{1ex} -\sum_{j\in \mathcal{N}_i}(\lambda_{ij}^{k-1}+\lambda_{ii}^{k-1})^T b_{i} + \frac{\rho_i^{k}}{2}\sum_{j\in \mathcal{N}_i\cup {i}}\|\tilde{x_{j}}^{k}-b_{i}\|^2   
\end{equation*}
\begin{equation*}
\lambda_{ij}^{k}=\lambda_{ij}^{k-1} + \rho_i^{k}(\tilde{x_{i}}^{k}-b_{j}^{k}) \hspace*{1ex} \forall (i,j) \in \mathcal{E}
\end{equation*}
\begin{equation*}
\lambda_{ii}^{k}=\lambda_{ii}^{k-1} + \rho_i^{k}(\tilde{x_{i}}^{k}-b_{i}^{k})    
\end{equation*}
Note that both the primal and dual updates are in a distributed manner, where only the information from neighbouring agents are required for each update.

The primal and dual residuals of the $i$th agent at the $k^{th}$ iteration i.e. $r_i^k$ and $s_i^k$ respectively, used to monitor the convergence of the algorithm, are defined as
$r_i^k=\|\tilde{x_i}^k-b_i^k\|_2+\sum_{j\in \mathcal{N}_i}\|\tilde{x_i}^k-b_j^k\|_2$
and
$s_i^k=\rho \|b_i^{k}-b_i^{k-1}\|_2$ respectively. 
For better convergence, the varying penalty parameter scheme described in \cite{boyd2011distributed} is followed:
\begin{equation*}
    \begin{split}
      \rho_i^{k+1}&=\nu \rho_i^k \hspace*{2.5ex} if \hspace*{1.5ex} \|r_i^k\|_2>\mu_1 \|s_i^k\|_2  \\
      &=\rho_i^k/\nu \hspace*{1.5ex} if \hspace*{1.5ex} \|s_i^k\|_2>\mu_2 \|r_i^k\|_2 \\
      &=\rho_i^k \hspace*{4ex} otherwise
    \end{split}
\end{equation*}
where $\nu>1$ and $\mu_1,\mu_2>1$ are parameters. 

In order to process new measurements and use the state estimate computed at the previous time step, in the sense of an observer, a time update step based on the system dynamics in \eqref{eqn:dyn_z} is performed i.e.
$$\tilde{z}_{Ti}(t)=\bar{A_i}\tilde{z_{i}}(t-1) + {N_i}\bar{y_{i}}(t)$$ where, $$\\
\bar{A_i}=\left[\begin{array}{cccc}
     A & 0 & \cdots{} & 0  \\
     G_{{1+\sum_{j=1}^{i-1}p_j}} & S & \cdots{} & 0 \\
     \vdots{}& &\ddots{} & \\
     G_{p_i+\sum_{j=1}^{i-1}p_j} & 0 & \cdots{} & S
\end{array}\right],{N_i}=\left[\begin{array}{c}\hspace*{-1.5ex}
0 \\ N_{{1+\sum_{j=1}^{i-1}p_j}} \\ \vdots{} \\ N_{p_i+\sum_{j=1}^{i-1}p_j}\hspace*{-1.5ex}
\end{array}\right]\\$$ 
and $\bar{y_{i}}(t)$ denotes the measurement vector at time $t$ for the $i$th observer. With the updated measurements, the optimization problem is again solved to get a better estimate of the state.  
The above steps are made precise in Algorithm \ref{algo_1} where the time update step forms the outer loop and the inner loop solves an optimization problem at each step to get a better estimate of the state in comparison to the previous time step. 
\begin{algorithm}
\caption{ADMM based Distributed Observer under Sparse Sensor attack}\label{algo_1}
\begin{algorithmic}[1]
\STATE {Initialize $\tau>0$ and collect measurements from time $t_0$ to $t_0+\tau-1$}
\STATE {Initialize $t=t_0+\tau$, $\lambda_{ii}^{0}, \lambda_{ij}^{0}, \rho_i^1, r_i^{t-1}, s_i^{t-1}$ and $\tilde{z_i}(t-1)$} 
For each observer $i$:
\WHILE{$r_i^{t-1}\geq \alpha$ \OR $s_i^{t-1} \geq \beta$} 
\STATE{\textbf{Time Update:}$\hspace*{1ex} \tilde{z}_{Ti}(t)=\bar{A_i}\tilde{z_{i}}(t-1) + {N_i}{y_{i}}(t)$}
\STATE Initialize k=1, {$b_i^0=\tilde{x}_{Ti}(t)$, $r_i^{t}=r_i^{t-1}$, $s_i^{t}=s_i^{t-1}$}
\WHILE{$r_i^{t}\geq (1-\nu)r_i^{t-1}$ \OR $s_i^{t} \geq (1-\nu)s_i^{t-1}$}
\STATE {$\tilde{z_{i}}^{k}= \underset{\tilde{z_{i}}=(\tilde{x_i},\tilde{E_i})}{\arg\min}\|\tilde{E_i}\|_1 + \sum_{j\in \mathcal{N}_i}(\lambda_{ij}^{k-1}+\lambda_{ii}^{k-1})^T \tilde{x_{i}} + \hspace*{3.5ex}\frac{\rho_i^k}{2}\sum_{j\in \mathcal{N}_i \cup {i}}\|\tilde{x_{i}}-b_{j}^{k-1}\|_2^2$
s.t. $Q_{i}\tilde{z_i}=Y_i(t)$}
\STATE {$b_{i}^{k}= \underset{b_{i}}{\arg\min}\hspace*{1ex} -\sum_{j\in \mathcal{N}_i}(\lambda_{ij}^{k-1}+\lambda_{ii}^{k-1})^T b_{i} + \hspace*{3.5ex}\frac{\rho_i^k}{2}\sum_{j\in \mathcal{N}_i\cup {i}}\|\tilde{x_{j}}^{k}-b_{i}\|_2^2$}
\STATE {$\lambda_{ij}^{k}=\lambda_{ij}^{k-1} + \rho_i^k(\tilde{x_{i}}^{k}-b_{j}^{k}) \hspace*{1ex} \forall (i,j) \in \mathcal{E}$}
\STATE {$\lambda_{ii}^{k}=\lambda_{ii}^{k-1} + \rho_i^k(\tilde{x_{i}}^{k}-b_{i}^{k})$}
\STATE {$r_i^t=\sum_{j\in \mathcal{N}_i\cup {i}}\|\tilde{x_i}^k-b_j^k\|_2$}
\STATE {$s_i^t=\rho_i^k \|b_i^k-b_i^{k-1}\|_2$}
\IF {$\|r_i^t\|_2 > \mu_1\|s_i^t\|_2$}
\STATE {$\rho_i^{k+1}=\nu \rho_i^k$}
\ELSIF{$\|s_i^t\|_2 > \mu_2\|r_i^t\|_2$}
\STATE {$\rho_i^{k+1}= \rho_i^k/\nu$}
\ELSIF{no update rules are triggered}
\STATE{$\rho_i^{k+1}= \rho_i^k$}
\ENDIF
\STATE {$k=k+1$} 
\ENDWHILE
\STATE{$\tilde{z_i}(t)=\tilde{z_{i}}^{k-1}, \rho_i^1=\rho_i^k, \lambda_{ii}^{0}=\lambda_{ii}^{k-1}$, $\lambda_{ij}^{0}=\lambda_{ij}^{k-1} ~\forall~ (i,j) \in \mathcal{E}$}
\STATE {$t=t+1$}
\ENDWHILE
\end{algorithmic}
\end{algorithm}

\textit{Stopping criterion}: According to \cite{boyd2011distributed}, the primal and dual residuals converge to zero as ADMM proceeds and when the primal and dual residuals are small, the objective suboptimality also must be small. Hence, suitable tolerances are chosen for  the primal and dual residuals to be used as stopping criteria for the algorithm. 

\begin{remark}
    At each time step, it is required to initialise the parameters $\lambda_{ii}$, $\lambda_{ij}$, $b_{i}$ and $\rho_i$. Since an optimization problem is solved at each time step to get better state estimates with updated measurements, we use the final values from the previous time step to initialize the parameters $\lambda_{ii}$, $\lambda_{ij}$ and $\rho_i$. The parameter $b_i$ introduced for consensus of state variable among observers is initialized to $\tilde{x}_{Ti}(t)$, the state estimate obtained after time update at the current time step.
\end{remark}

\begin{remark}
\textit{Convergence of Algorithm \ref{algo_1}} -
Algorithm 1 recursively solves the optimization problem in \eqref{dist_SBO} which is convex as both objective function and constraints are convex. We start with $\tau>0$ measurements collected from time $t=t_0$ to $t=t_0+\tau-1$ and use them to construct the matrices in the optimization problem in \eqref{dist_SBO}. Solving this problem gives the initial condition $x(t_0)$ and the attack vectors from time $t=t_0$ to $t=t_0+\tau-1$.  The primal and dual update steps in Section \ref{subsec:distr_observer} are the steps of a standard ADMM algorithm which provide the optimal solution. This along with residual monitoring constitute the inner loop of Algorithm 1. Since the ADMM algorithm converges (Theorem 22 in \cite{8931708}), the check condition in Step 6 of the algorithm would definitely be satisfied and hence the inner loop terminates. At each instance that the inner loop terminates, the time update (Step 4) is done. This update may increase disagreement among state values and cause the residual value to increase. Therefore the inner-loop, i.e. the ADMM algorithm, is performed again to reduce this increase in residual caused by the time-update. Since the check condition in Step 6 ensures that the residuals decrease across time, the outer loop terminates and the algorithm converges. 
\end{remark}

\begin{figure}
    \centering
   \includegraphics[width=0.33\textwidth]{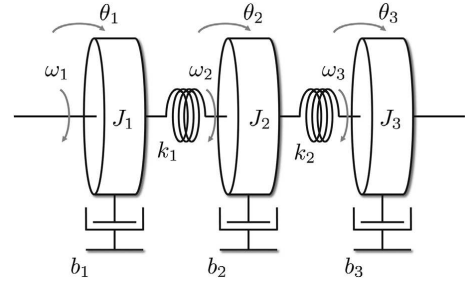}
    \caption{Three inertia system \cite{07654887318044adab44fdddb1abbca3}}
    \label{3 inertia system}
\end{figure}

\subsubsection*{Example}
The proposed Algorithm \ref{algo_1} is implemented for the three-inertia system shown in Figure \ref{3 inertia system} (considered in \cite{AN2019526},\cite{07654887318044adab44fdddb1abbca3},\cite{Zhang99}) which is $4$-sparse observable and has dynamics represented by the following continuous time state space equation:
\begin{eqnarray} \label{res_dynamics}
\dot x[t]=&A_cx[t]\nonumber \\
    y[t]=&C_cx[t]+a[t]
\end{eqnarray}
where $$A_c=\left[\begin{array}{cccccc}
\phantom{-}0&\phantom{-}1&\phantom{-}0&\phantom{-}0&\phantom{-}0&\phantom{-}0\\
-\frac{k_1}{J_1}&-\frac{b_1}{J_1}&\frac{k_1}{J_1}&\phantom{-}0&\phantom{-}0&\phantom{-}0\\
\phantom{-}0&\phantom{-}0&\phantom{-}0&\phantom{-}1&\phantom{-}0&\phantom{-}0\\
\phantom{-}\frac{k_1}{J_2}&\phantom{-}0&-\frac{k_1+k_2}{J_2}&-\frac{b_2}{J_2}&\phantom{-}\frac{k_2}{J_2}&\phantom{-}0\\
\phantom{-}0&\phantom{-}0&\phantom{-}0&\phantom{-}0&\phantom{-}0&\phantom{-}1\\
\phantom{-}0&\phantom{-}0&\phantom{-}\frac{k_2}{J_3}&\phantom{-}0&-\frac{k_2}{J_3}&-\frac{b_2}{J_3}\\
\end{array}\right]$$ $$C_c=\left[\begin{array}{cccccc}
1&0&\phantom{-}0&0&\phantom{-}0&0\\
0&0&\phantom{-}1&0&\phantom{-}0&0\\ \hline
0&0&\phantom{-}0&0&\phantom{-}1&0\\
1&0&-1&0&\phantom{-}0&0\\ \hline
1&0&\phantom{-}0&0&-1&0\\
0&0&\phantom{-}1&0&-1&0
\end{array}\right]=\left[\begin{array}{c}
C_1\\C_2\\C_3
\end{array}\right]$$ 
where $J_1=0.01\hspace*{0.4ex}kgm^2$, $J_2=0.02\hspace*{0.4ex}kgm^2$, $J_3=0.03\hspace*{0.4ex}kgm^2$ are the inertias of drive motor, middle body and load respectively, $K_1=K_2=1.4\hspace*{0.4ex}N/rad$ are the torsional stiffness of two shafts, $B_1=B_2=B_3=0.005\hspace*{0.4ex}N/(rad/s)$ are the mechanical damping of three inertias and $x=[\theta_1, \omega_1, \theta_2, \omega_2, \theta_3, \omega_3]$ where $\theta_1, \theta_2, \theta_3$ are the absolute angular positions of three inertias, $\omega_1, \omega_2, \omega_3$ are the speeds of three inertias and the output measurements are the three angular positions and the three relative angular positions. To obtain a discrete-time model, the system in (\ref{res_dynamics}) is discretized with a sampling period of $h=0.1s$.
The undirected communication graph of the 3 local observers is connected and the adjacency matrix of this graph is
$\mathcal{A}=\left[\begin{smallmatrix}
0&1&1\\
1&0&0\\
1&0&0
\end{smallmatrix}\right]$. The initial state of the system is taken as $x[0]=[0,0,0,0,0.9644,0]^T$. The third and fourth sensors are attacked and the attack vectors are generated at random. We have $\tau=3$ measurements collected at $t=0,1,2$ and the algorithm is initialised at $t=3$. Penalty parameter $\rho$ is initialised as 1 for all agents and $\nu, \mu_1, \mu_2$ are chosen as $10, 2.5$ and $1.1$ respectively. The stopping condition used is a tolerance of 0.1 for the primal and dual residuals. 
\begin{figure}
    \centering
   \includegraphics[width=0.5\textwidth]{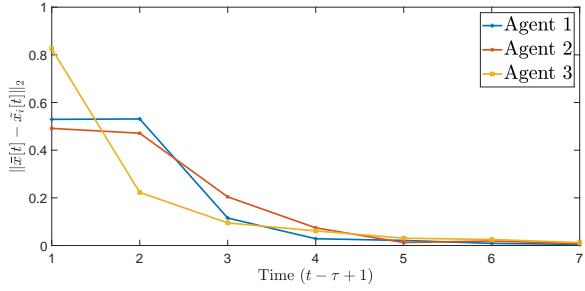}
    \caption{Evolution of the consensus error in the state estimates of all local observers across time}
    \label{dist_obs_cons}
\end{figure}

\begin{figure}
    \centering
   \includegraphics[width=0.5\textwidth]{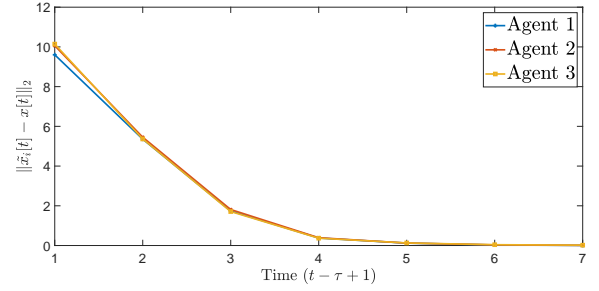}
    \caption{Evolution of error in the state estimate of all agents across time}
    \label{dist_obs_err}
\end{figure}
The comparison of consensus error associated with all the local observers is shown in Figure \ref{dist_obs_cons} and the evolution of error in state estimate of all the observers across time is shown in Figure \ref{dist_obs_err}. In this case, the average number of inner loop iterations the observer algorithm takes in a time step is 40.

We next adapt Algorithm 1 for the DSSE problem and the design of a centralized observer with sensor attacks and compare the performance with the existing algorithms in literature.
\subsubsection{Case 1 - Implementation as Distributed Secure State Estimator}

Algorithm \ref{algo_1} is modified to address the DSSE problem by executing the inner loop alone with loop guard as $r_i^{k}\leq \alpha$, $s_i^{k} \leq \beta$, for some small $\alpha,\beta$, instead of $r_i^{t}\leq (1-\nu) r_i^{t-1}$,  $s_i^{t} \leq (1-\nu) s_i^{t-1}$ to solve the optimization problem given in (\ref{dist_SBO}) i.e. to estimate the initial state $x[t_0]$ using a batch of measurements collected from time $t=t_0$ to $t=t_0+\tau-1$. 

\begin{figure}
    \centering
   \includegraphics[width=0.5\textwidth]{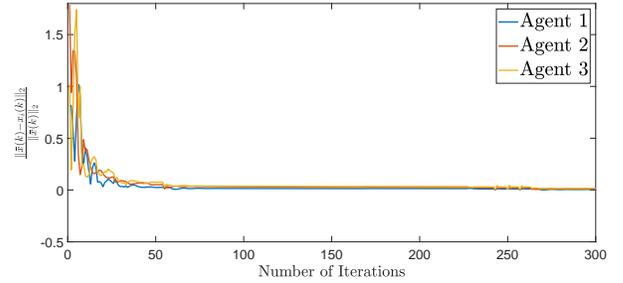}
    \caption{Evolution of consensus error in the agent state estimates in case of implementation as Distributed Secure State Estimator}
    \label{dist_static_cons}
\end{figure}

\begin{figure}
    \centering
   \includegraphics[width=0.5\textwidth]{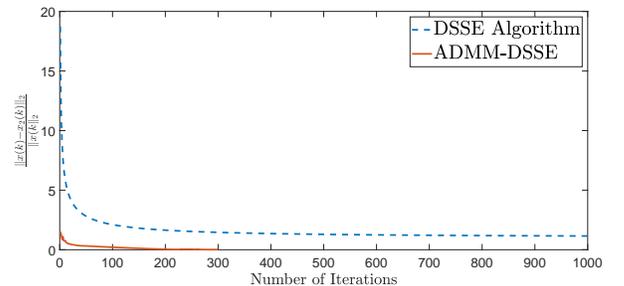}
    \caption{Comparison of the error in state estimate of agent 2
    }
    \label{dist_static_err}
\end{figure}

This idea is implemented and compared with the DSSE algorithm in \cite{07654887318044adab44fdddb1abbca3} for the system considered earlier. Since the DSSE algorithm assumes a sparse initial state, the initial state for the system is taken as $x[0]=[0,0.7196,0,0,0,0]^T$ and the third and sixth sensors are attacked. Penalty parameter $\rho$ is initialised as 1 for all agents and $\nu, \mu_1, \mu_2$ are chosen as $10, 2, 2$ respectively. Both the algorithms are run for 1000 iterations. Figure \ref{dist_static_cons} shows the comparison of consensus error associated with all the agents for our ADMM-based algorithm and Figure \ref{dist_static_err} shows the error in state estimate of agent 2 for both the algorithms. We observe that our algorithm converges faster.

\subsubsection{Case 2 - Implementation as Centralized Observer}
Since centralized observer under sparse sensor attacks exists in literature, we adapt our distributed observer in Algorithm 1 for the centralized case and compare its performance with the ETPL observer in \cite{7308014}.
\begin{algorithm}
\caption{ADMM based Centralized Observer}\label{algo_2}
\begin{algorithmic}[1]
\STATE {Initialize $\tau>0$ and collect measurements from time $t_0$ to $t_0+\tau-1$}
\STATE {Initialize $t=t_0+\tau$, $\lambda^{0},\rho>0, r^{t-1}$ and $\tilde{z}(t-1)$}
\WHILE{$r^{t-1}\geq \alpha$}
\STATE{\textbf{Time Update:}$\hspace*{1ex} \tilde{z}_{T}(t)=\bar{A}\tilde{z}(t-1) + {N}{y}(t)$}
\STATE Initialize $k=1$, { $r^{t}=r^{t-1}$} 
\WHILE{$r^{t}\geq (1-\nu)r^{t-1}$}
\STATE {$\tilde{z}^k=\underset{\tilde{z}=(\tilde{x},\tilde{E})}{\arg\min}\|\tilde{E}\|_1 + (\lambda^{k-1})^T(Q\tilde{z}-\bar{Y}) +\frac{\rho}{2}\|Q\tilde{z}-\bar{Y}\|_2^2$}
\STATE {$\lambda^k=\lambda^{k-1}+\rho(Q\tilde{z}^k-\bar{Y})$}
\STATE{$r^t=\|Q\tilde{z}^k-\bar{Y}\|_2$}
\STATE {$k=k+1$}
\ENDWHILE
\STATE{$\tilde{z}(t)=\tilde{z}^{k-1}, \lambda^{0}=\lambda^{k-1}$}
\STATE {$t=t+1$}
\ENDWHILE
\end{algorithmic}
\end{algorithm}
This is given as Algorithm \ref{algo_2}.

\begin{figure}
    \centering
    \includegraphics[width=0.5\textwidth]{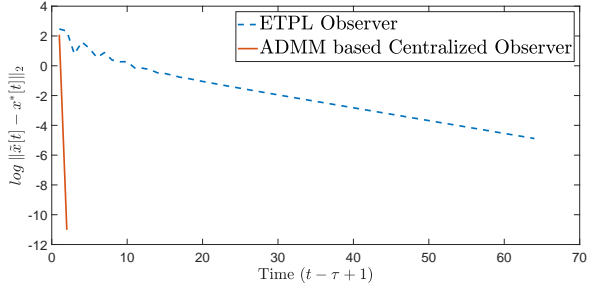}
    \caption{Comparison of the evolution of error (in logarithmic scale) in the system state estimate of both the algorithms. The stopping condition used is a tolerance of $10^{-5}$ for the primal residual.}
    \label{cent_error}
\end{figure}

We consider the system in (\ref{res_dynamics}) with initial state $x[0]=[0.5453,0.6888,0.1474,0.7776,0.3991,0.8983]^T$. The third and fourth sensors are attacked. Penalty parameter $\rho$ is chosen as 1 for Algorithm \ref{algo_2}. Figure \ref{cent_error} shows the evolution of the error in state estimate for both the algorithms. It can be observed that our algorithm converges significantly faster.

\section{Conclusion} \label{conc}
In this paper, we addressed the problem of designing a distributed observer for the state estimation of a discrete time LTI system under sparse sensor attack. An algorithm based on ADMM was proposed, for the update of each local observer, by which a network of observers were able to asymptotically estimate the system state using each of their limited, corrupt measurements and by communication with their neighbours. The proposed algorithm was adapted to solve the DSSE problem and also design a centralized observer. The performance was compared with other algorithms in literature for a well-studied system. As future work, adoption of event triggering techniques to reduce the number of communications among agents and distributed observer design under attack for directed networks could be of interest.

\bibliographystyle{IEEEtran}
\bibliography{references.bib}


%




\ifCLASSOPTIONcaptionsoff
  \newpage
\fi

\end{document}